\documentclass[prl,onecolumn,floatfix,a4paper,superscriptaddress]{revtex4}

\usepackage{bm,color,amsmath,txfonts}
\usepackage{graphicx}
\usepackage{siunitx}
\usepackage{subfigure}
\usepackage{verbatim}
\usepackage{dcolumn}% Align table columns on decimal point
\usepackage{bm}% bold math
\usepackage{epsf}
\usepackage{xcolor}
\usepackage{hyperref}
\usepackage{hhline}
\usepackage{float}
\usepackage{enumerate}
\usepackage{bbm}
\usepackage{lipsum}
\usepackage{mathrsfs}
\usepackage{ulem}

\begin{document}

\title{Comment on: ``Coherent perfect absorption: Zero reflection without linewidth suppression"}

\author{Rui-Chang Shen}
\affiliation{Department of Physics, The Chinese University of Hong Kong, Shatin, Hong Kong SAR, China}
\author{Jie Li}\thanks{jieli007@zju.edu.cn}
\affiliation{School of Physics, Zhejiang University, Hangzhou 310027, China}

\begin{abstract}
A recent paper, Phys. Rev. Research 8, 013261 (2026), claims that the polaromechanical normal-mode splitting (NMS) measured in Nat. Commun. 16, 5652 (2025) is not true based on their two results: $i$) there is no true splitting in the linear-scale spectrum; $ii$) the total or intrinsic decay rate of the cavity-magnon polariton, set by the imaginary part of the pole of the total output spectrum, remains unchanged under the coherent-perfect-absorption (CPA) condition. In this comment, we indicate that $i$) there is NMS in both the linear and logarithmic scales of our spectra in {\it a narrow frequency range} around the CPA frequency;  $ii$) the total decay rate defined via the {\it pole} of the spectrum cannot characterize the vanishing {\it effective} decay rate at the CPA frequency (known as the monochromaticity of the CPA), and thus this parameter is irrelevant to the NMS measured in our experiment in {\it a narrow frequency range} around the CPA frequency.  Consequently, their results above are either false or irrelevant, and thus cannot support their claim on the polaromechanical strong coupling measured in our experiment.

\end{abstract}

%\date{\today}
\maketitle

In what follows, we provide multiple evidences to show why the aforementioned two results of Ref.~\cite{JD} are either false or irrelevant, and how the CPA induces a significantly reduced {\it effective} decay rate around the CPA frequency, which leads to the polaromechanical strong coupling and the associated NMS measured in {\it a narrow frequency range} around the CPA frequency~\cite{Shen}. In the end, we indicate that the {\it effective} decay rate, defined via the imaginary part of the {\it zero} of the output spectrum~\cite{Hu}, can characterize the typical features of the CPA and is likely the correct way to interpret and understand the NMS measured in our experiment~\cite{Shen}. For convenience and consistency, the parameters and their definitions used in Secs. 1 and 2 are the same as in Ref.~\cite{Shen}.

\subsection{1. Polaromechanical normal-mode splitting in the linear-scale spectrum}
\label{sec1}

\begin{figure*}[t]
	\includegraphics[width=0.95\linewidth]{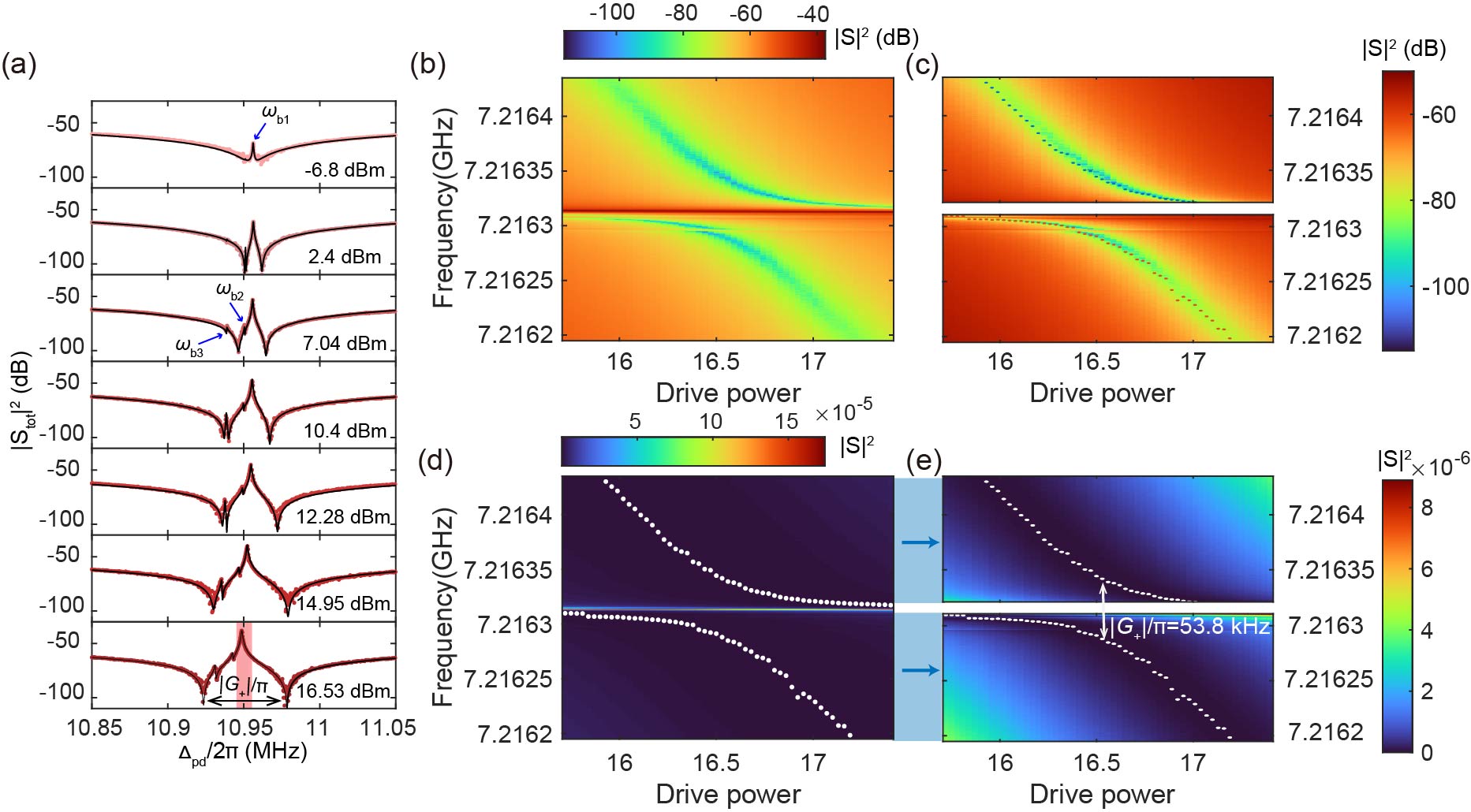}
	\caption{(a) Fig. 3b of Ref.~\cite{Shen}. (b) Logarithmic scale and (d) linear scale of the data of Fig. 4a in Ref.~\cite{Shen}. In the linear scale (d), the NMS is not clearly visible because the values of the red-shaded area in the bottom graph of (a), from 10.9458 to 10.9558 MHz, are several orders of magnitude larger than those in the rest part, which causes almost the entire area to be dark. However, the hidden NMS in (d) emerges when the data of the red-shaded area are not presented, yielding the blank area in (e). The same operation turns the logarithmic-scale (b) into (c). In (e), the two branches of the white dots  (measured data) denote the frequency shift of the two dips in (a), which clearly shows the splitting. }
	\label{fig1}
\end{figure*}

Here we show that the NMS is also present in our linear-scale spectrum. Figure~\ref{fig1}(a) is the Fig.~3b of Ref.~\cite{Shen}, and Figs.~\ref{fig1}(b) and~\ref{fig1}(d) are the logarithmic and linear scales of the data of Fig.~4a in Ref.~\cite{Shen}, respectively.  The dB-scale Fig.~\ref{fig1}(b) shows a clear NMS around the mechanical sideband at $2\pi \times 7.216313$ GHz, which equals to the CPA frequency $\omega_{\rm CPA}$. The NMS in the linear-scale Fig.~\ref{fig1}(d) is not clearly visible, because the values of the red-shaded area in the bottom graph of Fig.~\ref{fig1}(a) (from 10.9458 to 10.9558 MHz) are several orders of magnitude larger than those in the rest part, causing almost the entire Fig.~\ref{fig1}(d) to be dark, which buries the splitting.  However, the hidden NMS in Fig.~\ref{fig1}(d) emerges when the data of the red-shaded area are not presented, which results in the blank area in Fig.~\ref{fig1}(e).  The splitting of $53.8$ kHz in the spectrum approximately equals to twice the polariton-mechanics effective coupling strength $2|G_+|/(2\pi)=2\times 25.71 \approx 51.4$ kHz~\cite{Shen} extracted by fitting the spectrum using our theory, which is much larger than the {\it effective} decay rate of the polariton $\kappa_+/(2\pi) =0.45$ kHz and of the mechanical mode $\kappa_{\rm b,eff}/(2\pi) \approx 0.4$ kHz, quantitatively confirming that the polaromechanical system is in the strong-coupling regime.  

It should be noted that the polaromechanical NMS in Ref.~\cite{Shen} is analogous to the optomechanical NMS measured via the optomechanically induced transparency~\cite{JT}. Unlike the splitting measured in a relatively wide range~\cite{JT}, our NMS only occurs in {\it a narrow frequency range} around the CPA frequency, which, we will expound below, is a typical characteristic of the CPA.

\subsection{2. Vanishing polariton effective decay rate derived from non-Hermitian Hamiltonian with an effectively gained cavity mode}
\label{sec2}

As strictly derived in our Eq. (S16) in the supplementary information (SI) of Ref.~\cite{Shen}, the CPA, where $A_{j}^{\rm out} \,{=} \sqrt{2\kappa_j} A \,{-}\, a_{j}^{\rm in} \,{=}\, 0$ ($j=1,2$), can realize an {\it effective} gain of the cavity mode. The effectively gained cavity mode (via CPA) is {\it essential} for experimentally realizing the non-Hermitian PT-symmetric Hamiltonian of the cavity-magnon system and measuring the associated exceptional point (EP) in the experiment~\cite{Zhang}. This was clearly shown in Eqs.~(1)-(2) of Ref.~\cite{Zhang}.  We emphasize that the effective gain (of the cavity mode) exists not only mathematically but also physically -- it leads to {\it real physical effects}: without an effectively gained cavity mode, the PT-symmetric cavity-magnon system could not be constructed and the associated EP could not be measured in their experiment~\cite{Zhang}. 

Similarly, the effective gain of the cavity mode (via CPA) leads to real physical effects in our experiment~\cite{Shen}, i.e., the {\it effective} decay rates of the cavity-magnon polaritons vanish at the CPA frequency $\omega_{\rm CPA}$ (at which the CPA is realized, known as the monochromaticity of the CPA~\cite{CPA1,CPA2,CPA3}), $\kappa_{\pm} = (-\kappa_a'+ \kappa_m)/2 = 0$, when the cavity effective gain rate $\kappa_a'=\kappa_1 + \kappa_2 - \kappa_{\rm int}$ equals to the magnon loss rate $\kappa_m$, i.e., $\kappa_a'=\kappa_m$, and when the cavity and magnon modes are resonant. This was strictly derived in Eqs.~(S16)-(S19) of Ref.~\cite{Shen} using the non-Hermitian Hamiltonian of the cavity-magnon system. This is also straightforward to understand: the polariton is formed by the hybridization of the cavity mode and the magnon mode with equal weight for the resonant case, and balanced cavity gain rate and magnon loss rate naturally lead to a vanishing {\it effective} decay rate of the polariton, i.e., $\kappa_+ = (-\kappa_a'+ \kappa_m)/2 = 0$. Exactly, due to this significantly reduced effective decay rate $\kappa_+$ in {\it a narrow frequency range around the CPA frequency} (since the effective gain of the cavity is achieved via CPA only around $\omega_{\rm CPA}$), the polariton-mechanics strong coupling $G_+> \kappa_+, \kappa_{\rm b,eff}$ becomes possible, leading to the polaromechanical NMS measured in both the linear (Fig.1) and logarithmic (Fig.4a of Ref.~\cite{Shen}) scales. It should be noted that since the effective coupling strength $G_+$ is about tens of kHz~\cite{Shen}, the {\it effective} decay rate of the polariton $\kappa_+$ must be significantly reduced (around the CPA frequency), such that the strong coupling $G_+> \kappa_+$ is possible. Accordingly,  the NMS only occurs in {\it a narrow frequency range} around the CPA frequency, beyond which $\kappa_+>G_+$.  This means that the total decay rate via measuring the full-width-at-half-maximum (FWHM) of the broad spectral background~\cite{JD} can never capture this CPA-induced NMS within a narrow range. The statement ``{\it Claims of splitting require corroboration from both linear and logarithmic spectra over a sufficiently wide frequency range}" in Ref.~\cite{JD} clearly refers to the situation of the conventional NMS~\cite{Naka,Tang}, which is, however, not a characteristic of the CPA-induced splitting.

Similar NMS (strong coupling) was measured in an optomechanical system by introducing a feedback loop to achieve a significantly reduced cavity {\it effective} decay rate $\kappa_{\rm eff} < G$~\cite{DV}. Despite using different methods to get a significantly reduced effective decay rate, NMS occurs when the strong coupling condition $\kappa_{\rm eff} < G$~\cite{DV} or $\kappa_+ < G_+$~\cite{Shen} is satisfied (note: mechanical damping rate is typically much smaller). Besides, theory also indicates that coherent feedback, exploiting destructive interference similar to CPA, can also yield a greatly reduced {\it effective} decay rate of the cavity, which results in real physical effects, e.g., enhanced entanglement of two mechanical resonators~\cite{Li17}.

\subsection{3. Vanishing effective decay rate set by the imaginary part of the zero of the output spectrum}
\label{sec3}

Although Ref.~\cite{JD} shows that the total decay rate $\gamma$ (defined by the imaginary part of the {\it pole} of the total output spectrum $S_{\rm tot}$) remains unchanged under the CPA condition, which is correct and also shown in Ref.~\cite{Hu}, this parameter cannot characterize the CPA-induced significantly reduced {\it effective} decay rate {\it in a small range around the CPA frequency}, and thus is an irrelevant (and misleading) concept to interpreting the NMS measured in our experiment in both the linear and dB scales. A more relevant parameter is the effective decay rate $\gamma_{\rm eff}$, defined via the imaginary part of the {\it zero} of the output spectrum $S_{\rm tot}$  in Ref.~\cite{Hu} (see Ref.~\cite{Hu} for more details about the differences between the total/intrinsic decay rate $\gamma$ and the effective decay rate $\gamma_{\rm eff}$ under the CPA). The parameter $\gamma_{\rm eff}$ has a clear CPA feature, i.e., it vanishes {\it at the CPA frequency} under the CPA condition, while the total/intrinsic decay rate $\gamma$ remains unchanged.   Ref.~\cite{Hu} indicates that the effective decay rate $\gamma_{\rm eff}$ corresponds to the FWHM of the {\it inverse} spectrum $1/|S_{\rm tot}|^2$, while the conventional FWHM of the spectrum $|S_{\rm tot}|^2$ gives the the total decay rate $\gamma$.

For a directly coupled cavity-magnon system, $\gamma_{\rm eff}$ corresponds to our polariton effective decay rate $\kappa_+$ in Ref.~\cite{Shen}. Therefore, the significantly reduced $\gamma_{\rm eff}$ around the CPA frequency, rather than the unchanged $\gamma$ with/without CPA, should be adopted to compare with the effective coupling strength $G_+$ in our experiment in order to correctly interpret and understand the measured NMS in a narrow frequency range around the CPA frequency. Since Ref.~\cite{JD} focuses only on the study of the total decay rate $\gamma$ (defined via the {\it pole} of $S_{\rm tot}$) under the CPA, their results are irrelevant to the strong coupling (NMS) measured in our experiment, and thus their claim on the strong coupling of Ref.~\cite{Shen} is groundless and invalid.

\end{document}